\documentclass[twocolumn]{aastex701}

\usepackage{amsmath}
\usepackage{amssymb}
\usepackage{bm}
\usepackage{mathrsfs}
\usepackage{booktabs}
\usepackage{graphicx}
\usepackage[normalem]{ulem}

\providecommand{\bv}{}
\renewcommand{\bv}{{\bm v}}
\providecommand{\br}{}
\renewcommand{\br}{{\bm r}}
\newcommand{\I}{\mathrm{i}}
\newcommand{\avg}[1]{\left\langle #1\right\rangle}
\newcommand{\intdG}{\int \frac{d\bv}{4\pi}}
\newcommand{\Var}{\mathrm{Var}}

\newcommand{\oi}{[O\,{\sc i}]}
\newcommand{\caii}{[Ca\,{\sc ii}]}

\shorttitle{Intermittent Fast Flavor Conversion}
\shortauthors{Bao \& Addazi}

\begin{document}

\title{Intermittent Turbulence, Fast Flavor Conversion, and Observable Supernova Probes}

\author[0000-0000-0000-0000]{Yiwei Bao}
\email{sjtu0538015@sjtu.edu.cn}
\affiliation{Tsung-Dao Lee Institute, Shanghai Jiao Tong University, Shanghai 201210, China}
\affiliation{School of Physics and Astronomy, Shanghai Jiao Tong University, Shanghai 200240, China}

\author{Andrea Addazi}
\email{addazi@scu.edu.cn}
\affiliation{Center for Theoretical Physics, College of Physics Science and Technology, Sichuan University, 610065 Chengdu, China}
\affiliation{Istituto Nazionale di Fisica Nucleare, Laboratori Nazionali di Frascati, Frascati (Rome), Italy}

\begin{abstract}
Fast flavor conversion (FFC) in core-collapse supernovae is commonly studied in homogeneous backgrounds or with smooth stochastic closures for turbulence.  This work constructs an exact linear benchmark in which the matter-noise memory kernel is generated by a finite She--Leveque log-Poisson cascade.  Projection of a marginal FFC channel onto this kernel yields a causal Volterra equation whose non-Markovian memory closes into a finite local system.  The corresponding Laplace-space resolvent is rational, with one pole pair for each cascade level, so that the dispersion relation, characteristic polynomial, and time-domain solution can be verified analytically.  The benchmark is then coupled to a reduced realization model and a gain-region heating proxy.  For the updated intermittent choice $\delta\rho/\langle\rho\rangle=0.4$, $\bar\lambda/\mu=1$, and hence $\kappa_0=0.16$, the representative $N=2$, $r=2$ cascade gives $\sigma_{\rm int}^2=1.124$ and an intermittent conversion fraction $1-P_{\rm base}\simeq0.455$.  A weaker comparison normalization, $\kappa_0=0.05$, gives $1-P_{\rm base}\simeq0.324$.  The corresponding Mori-like heating ratios are $Q_{\rm int}/Q_{\rm hom}=1.060$ and $1.041$, whereas the Wang/Fornax-like ratios are $0.855$ and $0.899$.  Intermittency therefore sets the conversion fraction in the reduced model, while the neutrino spectral hierarchy determines the sign of the heating correction.
\end{abstract}

\keywords{Core-collapse supernovae (304) --- Neutrino astronomy (1100) --- Neutrino oscillations (1104) --- Supernova dynamics (1664) --- Astrostatistics (1882)}

\section{Introduction}

The delayed neutrino-heating mechanism remains the standard framework for core-collapse supernova explosions.  Neutrinos diffusing out of the proto-neutron star deposit a small fraction of their luminosity behind the stalled shock, and the outcome depends sensitively on the heating efficiency, accretion rate, turbulent pressure, and shock geometry \citep{BetheWilson1985,BurrowsGoshy1993,Janka2012,2013RvMP...85..245B,2016ARNPS..66..341J,2015PASA...32....9F,Burrows_2020,2022MNRAS.517..543W}.  Because many multidimensional models operate close to a critical condition rather than far above threshold, a few-percent change in the flavor-weighted heating budget can be dynamically relevant for marginal explosions \citep{BurrowsGoshy1993,MurphyBurrows2008,Janka2012,Burrows_2020,2022MNRAS.517..543W}.

Neutrino flavor conversion can therefore modify the heating problem in regimes close to criticality.  Supernova flavor evolution contains matter-enhanced conversion, collective effects, angular instabilities, and strong dependence on the local electron-lepton-number (ELN) distribution \citep{Wolfenstein_1978,Mikheyev_1985,duan_2010,2010ARNPS..60..569D,johns_2025}.  In the fast sector, the relevant scale is the neutrino self-interaction potential $\mu=\sqrt{2}G_F n_\nu$, and unstable modes are controlled primarily by the ELN angular structure rather than by the vacuum frequency \citep{2017PhRvL.118b1101I,2019PhRvD.100d3004A,2022PhRvL.128h1102D,johns_2025}.  Dispersion-relation analyses provide a systematic method for identifying local unstable branches and their growth rates \citep{2017PhRvL.118b1101I,2021PhRvL.126f1302B,2022PhRvL.128h1102D,2023PhRvL.131f1401E,2024PhRvL.133v1004F,Wang2025}.

The connection between an unstable root and the explosion engine is not unique.  The heating correction depends on how flavor conversion redistributes the $\nu_e$, $\bar\nu_e$, and $\nu_x$ spectra in the gain region \citep{Nagakura2023Heating,2023PhRvL.131f1401E,Wang2025}.  Depletion of electron flavor can reduce charged-current heating, whereas conversion from hotter heavy-lepton spectra can increase the charged-current energy deposition.  The sign of the correction is therefore model dependent.  This conclusion is supported by multi-angle FFC subgrid calculations coupled to multidimensional four-species Boltzmann neutrino radiation hydrodynamics, which find a bifurcated FFC impact: shock revival and explosion energy are enhanced in the lowest-mass progenitor but inhibited in higher-mass cases, with the mass accretion rate controlling whether spectral hardening or luminosity reduction dominates the heating response \citep{Akaho2026BifurcatedFFC}.  A benchmark based only on an averaged linear growth rate is therefore insufficient unless it also specifies how the growth history maps to survival probabilities and to a flavor-weighted heating proxy.

Turbulence adds a second layer of difficulty.  Stochastic matter effects in supernova flavor evolution have been studied for decades, including random MSW profiles, turbulence-induced stimulated transitions, and collective oscillations in fluctuating backgrounds \citep{Sawyer_1990,Loreti_1994,2011PhRvD..84j5034C,2010PhRvD..82l3004K,2014PhRvD..89g3022P,2015PhRvD..91b5001P,2014JCAP...11..030B,2021PhRvD.103d5014A}.  Most tractable closures, however, compress turbulence into a smooth correlation function.  That approximation is not guaranteed in the gain region, where convection, standing-accretion-shock-instability (SASI) motions, anisotropic accretion, and nonradial plumes are central to the dynamics \citep{Couch_2013,2015ApJ...799....5C,2015ApJ...808L..21C,Abdikamalov_2015,2017PhRvL.119x2702B,2019MNRAS.487.5304M,Burrows_2020,2022MNRAS.517..543W}.

Intermittency provides a further source of uncertainty.  Developed turbulence is not characterized by a single typical fluctuation; rare coherent structures dominate high-order moments, dissipation, and transport \citep{Kolmogorov_1941,SheLeveque1994,1995ApJ...438..763G,1995tlan.book.....F}.  If the flavor response is exponentially sensitive to local complex roots, rare cascade-tail episodes can contribute disproportionately to the effective growth history.  The log-Poisson cascade used below is an approximate closure rather than a substitute for multidimensional radiation-hydrodynamic or magnetohydrodynamic simulations.  Its purpose is to provide an analytically controlled representation of an intermittency hierarchy in the flavor-memory kernel.

This paper combines the analytic derivation, the reduced realization model, and the phenomenological heating estimate in a single formulation.  Section~\ref{sec:ffc} derives the reduced FFC channel.  Section~\ref{sec:cascade} constructs the log-Poisson kernel and identifies the entry point of $\delta\rho/\langle\rho\rangle$.  Section~\ref{sec:resolvent} gives the exact local closure and rational resolvent.  Section~\ref{sec:toy} defines the realization model and reports the two quantities used in the phenomenological analysis: the conversion fraction and the fractional heating change.

\section{Reduced Fast-Flavor Channel}
\label{sec:ffc}

A two-flavor mean-field description is adopted.  In the weak-interaction basis the density matrix can be written as
\begin{equation}
\varrho = \frac{f_{\nu_e}+f_{\nu_x}}{2}
+\frac{f_{\nu_e}-f_{\nu_x}}{2}
\begin{pmatrix}
s & S \\
S^* & -s
\end{pmatrix},
\label{eq:rho}
\end{equation}
where $S_\bv(t,\br)$ is the off-diagonal flavor coherence.  The ELN angular distribution is
\begin{equation}
G_\bv=\sqrt{2}G_F\int_0^\infty\frac{dE\,E^2}{2\pi^2}
\left[f_{\nu_e}(E,\bv)-f_{\bar\nu_e}(E,\bv)\right],
\label{eq:eln}
\end{equation}
and the local self-interaction scale is
\begin{equation}
\mu=\intdG\,G_\bv .
\label{eq:mu}
\end{equation}

The linearized homogeneous fast-mode equation has the standard form
\begin{equation}
\I(\partial_t+\bv\cdot\nabla_{\br})S_\bv
=\mathcal L_{\rm hom}[S]_\bv ,
\label{eq:linear_hom}
\end{equation}
where $\mathcal L_{\rm hom}$ contains the mean matter term, the ELN current, and the neutrino-neutrino angular coupling $(1-\bv\cdot\bv')G_{\bv'}$ \citep{duan_2010,2012JPhG...39c5201G,2017PhRvL.118b1101I,johns_2025}.  The matter potential is decomposed into a mean and a fluctuation,
\begin{equation}
\lambda(t,\br)=\bar\lambda+\delta\lambda(t,\br),
\label{eq:lambda_split}
\end{equation}
and average over a zero-mean turbulent ensemble.  A projection-operator or second-cumulant treatment gives a causal memory correction to the projected linear channel \citep{1960JChPh..33.1338Z,breuer2007theory,Loreti_1994,2010PhRvD..82l3004K,2014PhRvD..89g3022P}.  Schematically,
\begin{equation}
\I(\partial_t+\bv\cdot\nabla_{\br})\avg{S_\bv}
=\mathcal L_{\rm hom}\avg{S_\bv}
-\I\int_0^tdt'\,\mathcal K(t-t')\avg{S_\bv(t')}.
\label{eq:memory_general}
\end{equation}

For the analytic benchmark, the marginal ELN profile is taken to be
\begin{equation}
\frac{G_\bv}{\mu}=\frac{1+3\cos\theta}{2}.
\label{eq:marginal_eln}
\end{equation}
It has an ELN crossing at $\cos\theta=-1/3$ and is chosen so that the homogeneous channel is marginal at the selected point.  Projecting Eq.~(\ref{eq:memory_general}) onto this marginal eigenfunction gives a scalar amplitude $A(t)$ satisfying
\begin{equation}
\ddot A(t)+q^2A(t)
+2\I\int_0^tdt'\,\kappa(t-t')A(t')=0.
\label{eq:projected_dimful}
\end{equation}
The phase convention fixes the factor $2\I$; changing the convention complex conjugates the polynomial below but leaves $\max\Re s_j$ unchanged.  With $\tilde t=\mu t$, $\tilde q=q/\mu$, and $\tilde\kappa=\kappa/\mu^2$, Eq.~(\ref{eq:projected_dimful}) becomes
\begin{equation}
\partial_{\tilde t}^2 A+\tilde q^2A
+2\I\int_0^{\tilde t}d\tilde t'\,
\tilde\kappa(\tilde t-\tilde t')A(\tilde t')=0.
\label{eq:volterra}
\end{equation}
Dimensionless variables are used below unless stated otherwise.

\section{Intermittent Cascade Kernel}
\label{sec:cascade}

\subsection{Source of the Kernel}

The kernel is derived from the She--Leveque log-Poisson model of developed turbulence.  The structure functions obey
\begin{equation}
\avg{|\delta u_\ell|^p}\propto \ell^{\zeta_p},
\qquad
\zeta_p=\frac{p}{9}+C_0\left(1-\beta^{p/3}\right),
\label{eq:zeta_p}
\end{equation}
with $C_0=2$ and $\beta=2/3$ for filamentary dissipative structures \citep{SheLeveque1994,1995ApJ...438..763G,1995tlan.book.....F}.  Since the memory term is a two-point matter-fluctuation correlator, the relevant exponent is
\begin{equation}
\zeta_2=\frac{2}{9}+2\left[1-\left(\frac23\right)^{2/3}\right]\simeq0.696.
\label{eq:zeta2}
\end{equation}

The finite cascade is represented by a Prony sum,
\begin{equation}
\kappa(\tau)=\sum_{k=0}^{N}\Gamma_k
e^{-\tau/\tau_k}\cos(\Omega_k\tau),
\qquad \tau\ge0,
\label{eq:kernel_time}
\end{equation}
with cascade scalings
\begin{equation}
\tau_k=\tau_{\rm inj}r^{-2k/3},\qquad
\Omega_k=\Omega_{\rm inj}r^{2k/3},\qquad
\Gamma_k=\kappa_0 r^{-k\zeta_2}.
\label{eq:cascade_scalings}
\end{equation}
The turnover time follows $\tau_\ell\propto\ell^{2/3}$ for $\ell_k=\ell_{\rm inj}r^{-k}$, the frequency tracks $\tau_k^{-1}$, and the amplitude tracks the second-order structure function.  Thus the finite kernel is not fitted as a set of independent damped oscillators, but is a low-dimensional discretization of an intermittency hierarchy.

\subsection{Where \texorpdfstring{$\delta\rho/\langle\rho\rangle$}{drho/rho} Enters}

The matter fluctuation is $\delta\lambda=\sqrt{2}G_F\delta n_e$, so in the reduced dimensionless equation the base kernel amplitude scales as a squared fractional matter perturbation.  The adopted mapping is
\begin{equation}
\kappa_0=
\mathcal P_{\rm geom}
\left(\frac{\bar\lambda}{\mu}\right)^2
\left(\frac{\delta\rho}{\langle\rho\rangle}\right)^2 ,
\label{eq:kappa_density_mapping}
\end{equation}
where $\mathcal P_{\rm geom}$ is the projection factor from the local turbulent matter field onto the selected FFC eigenchannel.  Equation~(\ref{eq:kappa_density_mapping}) is the place where $\delta\rho/\langle\rho\rangle$ enters the flavor equation: it sets the normalization of $\Gamma_k$ in Eq.~(\ref{eq:cascade_scalings}), and hence the strength of the memory term in Eq.~(\ref{eq:volterra}).  The rest of the cascade prescription determines how this base coupling is distributed across levels.

For the fiducial intermittent benchmark,
\begin{equation}
\frac{\delta\rho}{\langle\rho\rangle}=0.4,\qquad
\frac{\bar\lambda}{\mu}=1,\qquad
\mathcal P_{\rm geom}=1,
\label{eq:canonical_density}
\end{equation}
so that
\begin{equation}
\kappa_0=0.16.
\label{eq:kappa0_updated}
\end{equation}
This value should be interpreted as a canonical local normalization rather than as a universal gain-region fluctuation amplitude.  A weaker comparison case, $\kappa_0=0.05$, is also retained; for $\bar\lambda/\mu=\mathcal P_{\rm geom}=1$ it corresponds to $\delta\rho/\langle\rho\rangle\simeq0.224$.  In simulation-calibrated applications, $\delta\rho/\langle\rho\rangle$, $\bar\lambda/\mu$, and $\mathcal P_{\rm geom}$ must be extracted along individual trajectories and correlated with the local ELN angular distribution.

\subsection{Intermittency Coordinate}

The log-Poisson mean at level $k$ is
\begin{equation}
\lambda_k=C_0\,\frac{k\ln r}{\ln(1/\beta)}.
\label{eq:lambda_k}
\end{equation}
If $m_k$ is Poisson distributed with mean $\lambda_k$, the multiplicative factor contains $\beta^{m_k}$ and
\begin{equation}
\Var(\ln W_k)=\lambda_k[\ln\beta]^2.
\label{eq:var_wk}
\end{equation}
The terminal variance
\begin{equation}
\sigma_{\rm int}^2=\Var(\ln W_N)
=C_0N\ln r\,\ln(1/\beta)
\label{eq:sigma_int}
\end{equation}
is used as the plotted intermittency strength.  The smooth limit is recovered by taking $\sigma_{\rm int}^2\to0$, equivalently $r\to1$ at fixed $N$, or by retaining only a single effective level.

The log-Poisson closure is approximate.  It retains scale-dependent intermittency and yields a solvable memory kernel, but compresses compressibility, anisotropy, magnetic stresses, matter-current fluctuations, shock curvature, neutrino-radiation feedback, and phase correlations into a small parameter set.  Full multidimensional radiation-hydrodynamic and MHD simulations are required to determine whether the assumed occupation statistics and coherence times are realized in the post-shock flow.

\section{Exact Resolvent and Local Closure}
\label{sec:resolvent}

For each cascade level define the quadratures
\begin{align}
X_k(t)&=\int_0^tdt'\,
e^{-(t-t')/\tau_k}\cos[\Omega_k(t-t')]A(t'),\\
Y_k(t)&=\int_0^tdt'\,
e^{-(t-t')/\tau_k}\sin[\Omega_k(t-t')]A(t').
\end{align}
Then
\begin{equation}
\int_0^tdt'\,\kappa(t-t')A(t')=\sum_{k=0}^{N}\Gamma_kX_k(t).
\label{eq:memory_xk}
\end{equation}
Leibniz differentiation closes the nonlocal Volterra equation into a local system:
\begin{align}
\dot A&=V,\\
\dot V&=-q^2A-2\I\sum_{k=0}^{N}\Gamma_kX_k,\\
\dot X_k&=A-\tau_k^{-1}X_k-\Omega_kY_k,\\
\dot Y_k&=\Omega_kX_k-\tau_k^{-1}Y_k.
\label{eq:closure}
\end{align}
The initial data are $A(0)=1$, $V(0)=0$, and $X_k(0)=Y_k(0)=0$.
Writing $\Psi=(A,V,X_0,Y_0,\ldots,X_N,Y_N)^T$ gives $\dot\Psi=M_{\rm casc}\Psi$.  This construction is an exact closure for the finite kernel and does not invoke a Markov approximation.

With the Laplace convention $\hat f(s)=\int_0^\infty e^{-st}f(t)\,dt$,
\begin{equation}
\hat\kappa(s)=
\sum_{k=0}^{N}\Gamma_k
\frac{s+\tau_k^{-1}}{(s+\tau_k^{-1})^2+\Omega_k^2}.
\label{eq:kappahat}
\end{equation}
The amplitude resolvent is
\begin{equation}
\hat A(s)=
\frac{s}{s^2+q^2+2\I\sum_{k=0}^{N}\Gamma_k(s+\tau_k^{-1})/D_k(s)},
\label{eq:Ahat}
\end{equation}
where
\begin{equation}
D_k(s)=\left(s+\tau_k^{-1}\right)^2+\Omega_k^2 .
\label{eq:Dk}
\end{equation}
Multiplying by $\prod_jD_j(s)$ gives the characteristic polynomial
\begin{align}
P_{2N+4}(s)={}&
(s^2+q^2)\prod_{j=0}^{N}D_j(s)
\nonumber\\
&+2\I\sum_{k=0}^{N}\Gamma_k(s+\tau_k^{-1})
\prod_{\substack{j=0\\j\ne k}}^{N}D_j(s).
\label{eq:poly}
\end{align}
The local matrix gives the same polynomial:
\begin{equation}
\det(sI-M_{\rm casc})=P_{2N+4}(s).
\label{eq:matrix_poly}
\end{equation}
For simple roots $s_j$,
\begin{equation}
A(t)=\sum_{j=1}^{2N+4}c_je^{s_jt},
\qquad
c_j=\operatorname*{Res}_{s=s_j}\hat A(s),
\label{eq:time_solution}
\end{equation}
and the linear cascade growth rate is
\begin{equation}
\gamma_{\rm casc}=\max_j\Re s_j .
\label{eq:gamma_casc}
\end{equation}

For the representative $N=2$, $r=2$ case, Eq.~(\ref{eq:poly}) is eighth order.  The numerical matrix-polynomial identity gives a maximum difference of $3.8\times10^{-15}$.  For $\kappa_0=0.16$, the level amplitudes are $\Gamma_0=0.1600$, $\Gamma_1=0.0988$, and $\Gamma_2=0.0610$; the full cascade gives $\gamma_{\rm casc}=0.5229$, while the injection-only cascade gives $0.3919$.  For $\kappa_0=0.05$, the corresponding level amplitudes are $0.0500$, $0.0309$, and $0.0191$, and the full cascade growth is $\gamma_{\rm casc}=0.2502$.  The cascade tail therefore modifies the pole spectrum and the growth history rather than simply perturbing a single smooth-memory branch.  The phenomenological observables reported below are the conversion fraction $1-P_{\rm base}$ and the fractional heating change $\Delta Q/Q_{\rm hom}$.

\section{Reduced Realizations and Heating Map}
\label{sec:toy}

The exact resolvent describes the ensemble-averaged reduced amplitude.  A survival probability, however, is sampled along individual trajectories.  A reduced realization model is therefore introduced to connect the cascade benchmark to a bounded flavor-conversion proxy.  The no-turbulence FFC baseline is a two-angle ELN crossing,
\begin{align}
(v_1,v_2)&=(-0.45,0.85),&
(g_1,g_2)&=(0.65,-0.50),\nonumber\\
\mu_{\rm hom}&=0.2 .
\label{eq:hom_toy_data}
\end{align}
For real $k_z$,
\begin{equation}
\mathsf M_{ij}(k_z)=\delta_{ij}v_ik_z-\mu_{\rm hom}(1-v_iv_j)g_j,
\label{eq:hom_toy_matrix}
\end{equation}
with normal modes $S_i=Q_i\exp[-\I(\omega t-k_zz)]$.  The homogeneous growth rate is
\begin{equation}
\gamma_{\rm hom}=\max_{k_z,j}\Im\omega_j(k_z),
\qquad \omega_j(k_z)\in{\rm spec}\,\mathsf M(k_z).
\label{eq:gamma_hom}
\end{equation}

For each cascade level $k\ge1$ the log-Poisson statistics define an on/off rare-event process $\chi_k(t)\in\{0,1\}$.  The transition rates are chosen as
\begin{equation}
p_k=1-e^{-\lambda_k},\qquad
r_k^{\rm off}=\tau_k^{-1},\qquad
r_k^{\rm on}=(e^{\lambda_k}-1)\tau_k^{-1},
\label{eq:markov_rates}
\end{equation}
so that the stationary occupation is $p_k$.  The per-level excess growth is defined by removing one cascade level at a time from the exact polynomial,
\begin{equation}
\Delta\gamma_k=\gamma_{\rm full}-\gamma_{-k}.
\label{eq:delta_gamma}
\end{equation}
The realization-level amplitude evolves as
\begin{equation}
\frac{d}{dt}\ln A_{\rm toy}(t)
=\gamma_{\rm hom}+\sum_{k=1}^{N}\Delta\gamma_k\chi_k(t).
\label{eq:Atoy}
\end{equation}
The corresponding smooth closure replaces $\chi_k(t)$ by $p_k$:
\begin{equation}
A_{\rm sm}(t)=
\exp\left[\left(\gamma_{\rm hom}+\sum_{k=1}^{N}p_k\Delta\gamma_k\right)t\right].
\label{eq:Asmooth}
\end{equation}

The survival proxy is a bounded monotonic map from above-threshold integrated coherence to conversion:
\begin{align}
P_{\rm base}
&=1-\mathcal C_{\max}
\left[1-\exp\left(-\frac{\mathcal I[A]}{S_{\rm sat}}\right)\right],\\
\mathcal I[A]&=\int_0^T dt\,
\max\left(A^2(t)-A_{\rm thr}^2,0\right).
\label{eq:Pbase}
\end{align}
The numerical parameters are $T=5$, $A_{\rm thr}=2$, $\mathcal C_{\max}=0.5$, and $S_{\rm sat}=5$.  At $\sigma_{\rm int}^2=0$ all $p_k$ vanish and the model returns the homogeneous baseline $P_{\rm base}=0.9762$, or conversion fraction $1-P_{\rm base}=0.0238$.  At the representative intermittent point, the mean survival is about $0.545$ for $\kappa_0=0.16$ and $0.676$ for $\kappa_0=0.05$, corresponding to conversion fractions $0.455$ and $0.324$.  These conversion fractions are listed in Table~\ref{tab:toy_outputs}.

The charged-current gain-region heating proxy keeps the standard luminosity and squared-energy scaling \citep{Janka2012,Nagakura2023Heating},
\begin{equation}
Q\propto
Y_n\Phi_{\nu_e}^{\rm eff}
+Y_p\Phi_{\bar\nu_e}^{\rm eff},
\qquad
Y_n=0.6,\quad Y_p=0.4 ,
\label{eq:Qproxy}
\end{equation}
with
\begin{align}
\Phi_{\nu_e}^{\rm eff}
&=P_{ee}L_{\nu_e}\avg{E_{\nu_e}}^2
+(1-P_{ee})L_{\nu_x}\avg{E_{\nu_x}}^2,\\
\Phi_{\bar\nu_e}^{\rm eff}
&=\bar P_{ee}L_{\bar\nu_e}\avg{E_{\bar\nu_e}}^2
+(1-\bar P_{ee})L_{\nu_x}\avg{E_{\nu_x}}^2.
\label{eq:phi_eff}
\end{align}
The split survival prescription is
\begin{align}
P_{ee}&=1-\eta_{\nu_e}(1-P_{\rm base}),&
\bar P_{ee}&=P_{\rm base},\nonumber\\
\eta_{\nu_e}&=\min(L_{\bar\nu_e}/L_{\nu_e},1).
\label{eq:split_survival}
\end{align}
This prescription is a reduced heating map rather than a transport solution.  It identifies which spectral combinations lead additional conversion to increase or decrease the gain-region heating proxy.  Differentiating with respect to the conversion fraction gives
\begin{align}
\Xi={}&Y_n\eta_{\nu_e}
\left[L_{\nu_x}\avg{E_{\nu_x}}^2
-L_{\nu_e}\avg{E_{\nu_e}}^2\right]\nonumber\\
&+Y_p
\left[L_{\nu_x}\avg{E_{\nu_x}}^2
-L_{\bar\nu_e}\avg{E_{\bar\nu_e}}^2\right].
\label{eq:Xi}
\end{align}
Positive $\Xi$ corresponds to an increase in the heating proxy under additional conversion.

\begin{figure*}[t]
\centering
\includegraphics[width=0.95\textwidth]{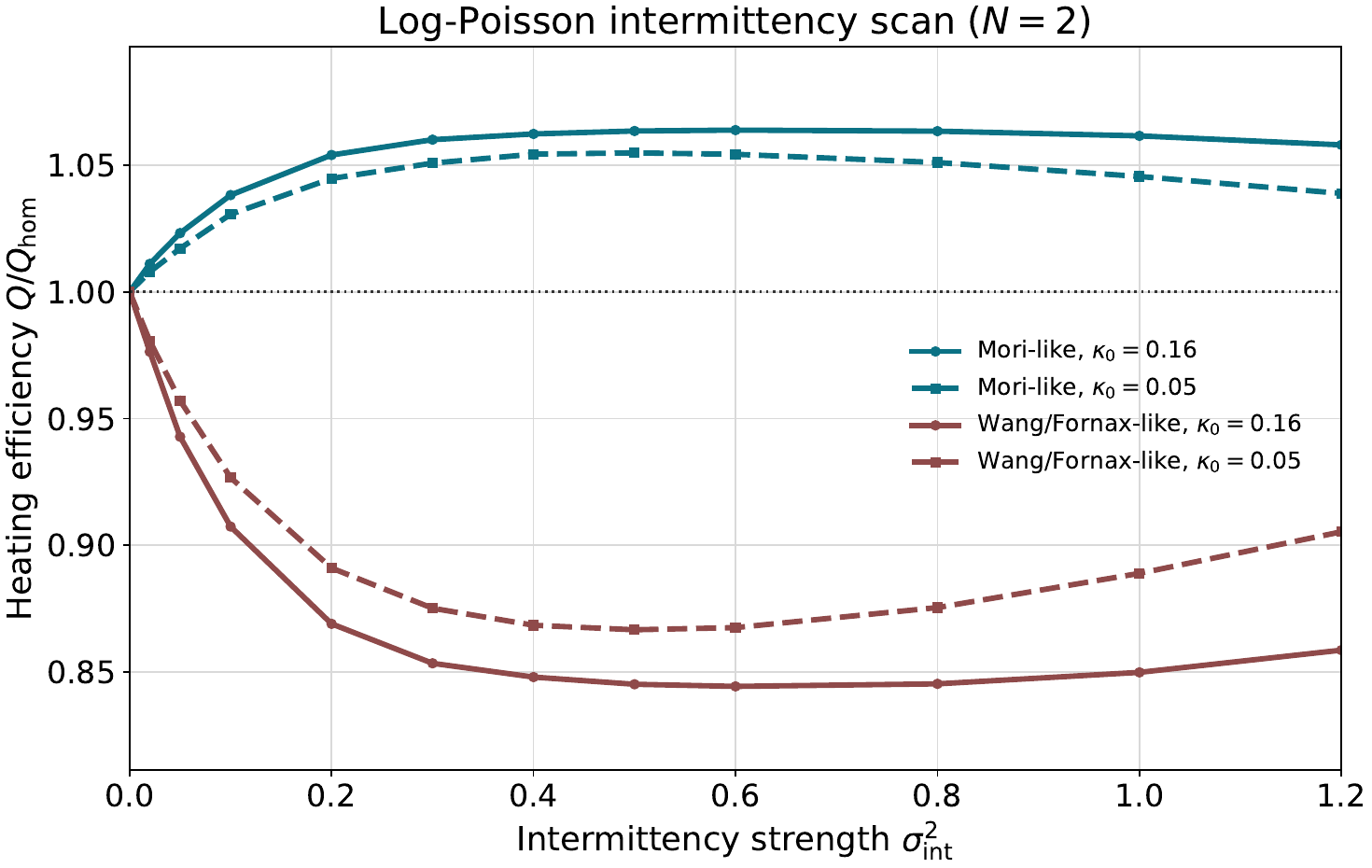}
\caption{Reduced gain-region heating efficiency normalized to the homogeneous no-turbulence baseline, $Q/Q_{\rm hom}$, as a function of log-Poisson intermittency strength $\sigma_{\rm int}^2=\Var(\ln W_N)$ for a fixed $N=2$ cascade.  All curves are shown on the same axes.  Colors distinguish the Mori et al. spectra \citep{2023PhRvD.108f3027M} from the Wang/Fornax spectra \citep{2022MNRAS.517..543W}; line styles distinguish the updated $\kappa_0=0.16$ case implied by $\delta\rho/\langle\rho\rangle=0.4$, $\bar\lambda/\mu=1$, and $\mathcal P_{\rm geom}=1$ from the weaker $\kappa_0=0.05$ comparison.  At the representative $r=2$ point, $\sigma_{\rm int}^2=1.124$; the Mori-like ratios are $Q_{\rm int}/Q_{\rm hom}=1.0596$ and $1.0415$, while the Wang/Fornax-like ratios are $0.8547$ and $0.8989$ for $\kappa_0=0.16$ and $0.05$, respectively.}
\label{fig:heating_scan}
\end{figure*}

The two spectral inputs are normalized to $L_{\nu_e}=1$.  The Mori-like benchmark adopts $(L_{\nu_e},L_{\bar\nu_e},L_{\nu_x})=(1,1.006908,0.642106)$ and $(\avg{E_{\nu_e}},\avg{E_{\bar\nu_e}},\avg{E_{\nu_x}})=(12.98,15.85,18.94)\,\mathrm{MeV}$.  The Wang/Fornax-like benchmark adopts $(1,0.929028,0.459257)$ and $(12.52,14.72,15.78)\,\mathrm{MeV}$.
For the representative cascade, Fig.~\ref{fig:heating_scan} gives
\begin{align}
\left(Q_{\rm int}/Q_{\rm hom}\right)_{\rm Mori}^{0.16}&=1.0596,\nonumber\\
\left(Q_{\rm int}/Q_{\rm hom}\right)_{\rm Mori}^{0.05}&=1.0415,\nonumber\\
\left(Q_{\rm int}/Q_{\rm hom}\right)_{\rm Wang}^{0.16}&=0.8547,\nonumber\\
\left(Q_{\rm int}/Q_{\rm hom}\right)_{\rm Wang}^{0.05}&=0.8989.
\label{eq:heating_results}
\end{align}
The smooth and intermittent means are nearly identical at this representative point because the selected $r=2$ cascade has high activation probabilities.  At weaker intermittency, however, the median, mean, and smooth closure can differ substantially because many trajectories avoid the strongest cascade events.

\begin{table}[t]
\centering
\caption{Reduced-model outputs for the representative intermittent benchmark.}
\label{tab:toy_outputs}
\begin{tabular}{lccc}
\toprule
$\kappa_0$ & Spectra & $\mathcal C_{\rm conv}$ & $\Delta Q/Q_{\rm hom}$ \\
\midrule
0.16 & Mori & 0.4546 & $+5.96\%$ \\
0.16 & Wang/Fornax & 0.4547 & $-14.53\%$ \\
0.05 & Mori & 0.3238 & $+4.15\%$ \\
0.05 & Wang/Fornax & 0.3237 & $-10.11\%$ \\
\bottomrule
\end{tabular}
\end{table}

\section{Observable Proxies}
\label{sec:obs}

The phenomenological analysis is restricted to two quantities that are directly connected to the reduced realization model.  The first is the conversion fraction,
\begin{equation}
\mathcal C_{\rm conv}\equiv 1-P_{\rm base},
\label{eq:Cconv}
\end{equation}
and the fractional heating change,
\begin{equation}
\frac{\Delta Q}{Q_{\rm hom}}\equiv
\frac{Q_{\rm int}}{Q_{\rm hom}}-1 .
\label{eq:deltaQ}
\end{equation}
These quantities can be evaluated without introducing additional electromagnetic modeling assumptions.  A Galactic neutrino signal would constrain $\mathcal C_{\rm conv}$ through flavor-dependent event rates and spectra, while the heating proxy determines whether the same conversion increases or decreases the gain-region energy deposition.  The sign of $\Delta Q$ is not fixed by intermittency alone; it is fixed by the underlying luminosity and mean-energy hierarchy through Eq.~(\ref{eq:Xi}).

\section{Discussion}

The principal theoretical result is that an intermittent cascade changes the structure of the FFC memory problem.  A smooth closure has one effective memory scale, whereas the finite log-Poisson kernel has a pole pair for each cascade level.  The reduced Volterra problem remains exactly solvable, so the result does not depend on a stochastic Monte Carlo approximation.  The exact solution also distinguishes the ensemble-averaged growth rate from a typical survival probability: intermittency can shift the dominant averaged pole while many individual trajectories remain closer to the homogeneous baseline.

The principal phenomenological result is that the sign of the heating correction is controlled by the neutrino spectra.  With $\delta\rho/\langle\rho\rangle=0.4$ and canonical $\bar\lambda/\mu=1$, the updated $\kappa_0=0.16$ realization model gives a few-percent enhancement for the Mori-like spectral hierarchy but a suppression for the Wang/Fornax-like hierarchy.  Reducing the kernel normalization to the weaker $\kappa_0=0.05$ case decreases the magnitude of both effects without changing their signs.  This behavior follows directly from Eq.~(\ref{eq:Xi}): intermittency supplies conversion, while the luminosity and mean-energy hierarchy determines whether conversion increases or decreases the heating proxy.  An analogous bifurcation appears in the multi-angle radiation-hydrodynamic study of \citet{Akaho2026BifurcatedFFC}, where FFC can either assist or inhibit the explosion depending on the accretion state and the competition between spectral hardening and luminosity reduction.

The most informative electromagnetic probes are therefore not broadband luminosities alone, but species-resolved line profiles that separate inner-ejecta structure from CSM or shell interaction.  If intermittent neutrino-driven plumes leave a lasting imprint on the ejecta, the cleaner late-time tracers are expected to be inner nucleosynthetic lines such as \oi\ $\lambda\lambda6300,6364$ and Fe/Co/Ni nebular features; relevant diagnostics include persistent velocity-space asymmetry, clumpiness, and heavy-tailed profile structure after corrections for doublets and radiative-transfer effects \citep{FranssonChevalier1989,Jerkstrand2012,Maeda2008,Milisavljevic2010}.  By contrast, H$\alpha$, \caii\ $\lambda\lambda7291,7323$, ultraviolet resonance features, and dust-sensitive line wings are more susceptible to CSM interaction, swept-up shells, or asymmetric external material.  SN~2023ixf illustrates this separation: early spectra and polarization indicate asymmetric CSM, whereas later polarimetry and nebular modeling require an additional inner-ejecta or helium-core asymmetry \citep{Ferrari2024ixfNebular,Singh2024ixfAsphericities,Shrestha2025ixfPolarimetry}.  Element-line diagnostics are therefore a natural observational extension of the present reduced model, but dedicated radiation-hydrodynamic and radiative-transfer calculations are required before they can be mapped quantitatively to $\mathcal C_{\rm conv}$ or $\Delta Q/Q_{\rm hom}$.

Several limitations follow from the reduced formulation.  The marginal ELN profile in Eq.~(\ref{eq:marginal_eln}) is a benchmark rather than a survey of realistic angular distributions.  The log-Poisson kernel is a controlled approximation to intermittent turbulence rather than a full MHD model.  The survival map and heating proxy are intentionally minimal.  A realistic calculation must extract $\delta\rho/\langle\rho\rangle$, matter-current fluctuations, correlation times, anisotropy, and ELN distributions from multidimensional simulations, and then evolve flavor conversion consistently enough to assess feedback on transport and heating.  Magnetic stresses may be relevant in some progenitors and rotation states, so MHD simulations are part of the required validation path.

The observational interpretation is consequently limited to diagnostic guidance.  In the present reduced treatment the controlled outputs are $\mathcal C_{\rm conv}$ and $\Delta Q/Q_{\rm hom}$.  The element-line probes discussed above are proposed diagnostics rather than additional quantitative predictions of the model.

\section{Conclusions}

This work formulates intermittent turbulent FFC with an exact finite-dimensional closure of a log-Poisson memory kernel.  The resulting amplitude equation has a rational resolvent and a characteristic polynomial identical to the local matrix determinant.  The density fluctuation enters through the base kernel normalization $\kappa_0$, and the fiducial choice $\delta\rho/\langle\rho\rangle=0.4$ gives $\kappa_0=0.16$ for canonical $\bar\lambda/\mu=1$.  The weaker normalization $\kappa_0=0.05$ is retained as a comparison case.

In the reduced heating map, the representative cascade has $\sigma_{\rm int}^2=1.124$.  For $\kappa_0=0.16$ it gives $Q_{\rm int}/Q_{\rm hom}=1.0596$ for Mori-like spectra but $0.8547$ for Wang/Fornax-like spectra; for $\kappa_0=0.05$ the corresponding ratios are $1.0415$ and $0.8989$.  Intermittency controls the trajectory distribution and conversion strength, while the neutrino spectral hierarchy controls the sign of the heating correction.

Future applications require kernels calibrated from multidimensional radiation-hydrodynamic and MHD simulations, including local ELN angular distributions.  The most direct quantities to carry forward are the flavor-conversion fraction and the associated heating change; indirect electromagnetic diagnostics require a separate radiation-hydrodynamic and radiative-transfer modeling layer.

\begin{acknowledgments}
We thank Thierry Foglizzo for insightful discussions.  This work is supported by the Fundamental Research Funds for the Central Universities under No. 020114380057, and by K. C. Wong Educational Foundation.  A.A. is supported by the National Science Foundation of China (NSFC) through grant No. 12350410358; the Talent Scientific Research Program of College of Physics, Sichuan University, Grant No. 1082204112427; the Fostering Program in Disciplines Possessing Novel Features for Natural Science of Sichuan University, Grant No. 2020SCUNL209; and the 1000 Talent program of Sichuan province 2021.
\end{acknowledgments}

\software{Python, NumPy, SciPy, matplotlib, Mathematica}

\bibliographystyle{aasjournalv7}
\bibliography{ffc}

\end{document}